\documentclass[aps,onecolumn,10pt]{revtex4}
\usepackage{amsmath}
\usepackage{amssymb}
\usepackage{hyperref}
\usepackage{graphicx}
\usepackage{epsfig}
\usepackage{float}
\usepackage{slashed} 
\hypersetup{colorlinks=true}
\begin{document}

\title{Is dark matter fact or fantasy? -- clues from the data}

\author{Philip D. Mannheim \\}
\affiliation{Department of Physics, University of Connecticut, Storrs, CT 06269, USA \\
philip.mannheim@uconn.edu}
\smallskip
\date{May 5, 2019}

\begin{abstract}
We discuss arguments both in favor of and against dark matter. With the repeated failure of experiment to date to detect dark matter we discuss what could be done instead, and to this end look for clues in the data themselves. We identify various regularities in galactic rotation curve data that correlate the total gravitational potential with luminous matter rather than dark matter. We identify a contribution to galactic rotation curves coming from the rest of the visible Universe, and suggest that dark matter is just an attempt to describe this global effect in terms of standard local Newtonian gravity within galaxies. Thus the missing mass is not missing at all -- it has been hiding in plain sight all along as the rest of the visible mass in the Universe.
 \end{abstract}

\maketitle
\smallskip
\noindent
\centerline{Essay written for the Gravity Research Foundation 2019 Awards for Essays on Gravitation}

\bigskip

\section{The Hundred Year Dark Matter Problem}
\label{S1}

In one form or another the dark matter problem has been with us for close to one hundred years. Remarkably, almost as soon as it was realized that the Milky Way was a galaxy, Oort found in the 1920s that the velocities of stars perpendicular to the plane of the galaxy had a missing mass problem. Then, almost as soon as it was realized that there were other galaxies, in the 1930s Zwicky and Smith found that the velocity dispersions of galaxies in a cluster of galaxies also had a missing mass problem. In the 1970s and 1980s $HI$ radio studies of Freeman \cite{Freeman1970} and of Roberts and Whitehurst \cite{Roberts1975} and $HII$ optical studies of Rubin, Ford and  Thonnard \cite {Rubin1980} showed that the measured rotational velocities in the outskirts of spiral galaxies greatly exceeded the luminous Newtonian expectations. Then the dam broke, with missing mass problems being found all the way up to cosmology, with cosmology even yielding an additional problem to boot -- dark energy.

Possible astrophysical options for dark matter, namely that dark matter was too faint to be detected or that it was in a non-luminous astrophysical form such as black holes or white dwarfs (machos), have been ruled out by improvements in telescopes and by gravitational lensing off the Magellanic clouds. Particle physics options for dark matter, namely particles that are intrinsically unable to emit light at all (supersymmetric particles, axions -- collectively wimps) have yet to be detected, with extensive, decades long accelerator and underground searches having not yet been able to find any wimps. Now while wimps are not ruled out, they are far from being ruled in, and there is even a tension for supersymmetry, the community's preferred form of dark matter,  since no superparticles in the Higgs boson mass region have been found, particles that were thought would solve the elementary Higgs boson hierarchy problem. 

Additionally, no solution to the dark energy/cosmological constant problem has been found either in the twenty or so years now since the discovery of the accelerating Universe. Now this problem actually predates the accelerating Universe. In fact Einstein introduced the cosmological constant in the 1920s, so this problem is also one hundred years old. And if $\Omega_{\Lambda}$ really is of order $10^{60}$ (its particle physics expectation) then none of the CMB (cosmic microwave background) background or fluctuation tests would even be remotely successful. 

And lurking behind all this is the hope that the standard Newton-Einstein classical gravity expectations are not destroyed by quantum mechanics. (That gravity knows about large scale quantum effects is evidenced by the stabilizing effect of the Pauli degeneracy of electrons in a white dwarf, and by the intrinsically quantum mechanical nature of the CMB black body radiation spectrum.) With Einstein gravity having been developed by Einstein in 1915, with quantum field theory have been developed by Dirac, Feynman, Schwinger, Tomonaga, and Dyson in the 1930s and 1940s, and with Einstein gravity not being renormalizable, the quantum gravity problem is more then 70 years old.

In this paper we will address the dark matter problem, and we will identify a shortcoming in the reasoning that leads us to dark matter in the first place -- namely taking the standard Newton-Einstein theory to be necessary to give Newton's Law of Gravity and the Schwarzschild solution rather than sufficient. To see what is to be required astrophysically, we shall look for clues in the data.  The clues that we present will not exclude dark matter but they will challenge it, and will not point in the direction of dark matter. We will focus primarily on galactic rotation curves of spirals since no theory is needed   -- just orbits. Some typical ones are shown in Fig. 1.
\begin{figure}[H]
\epsfig{file=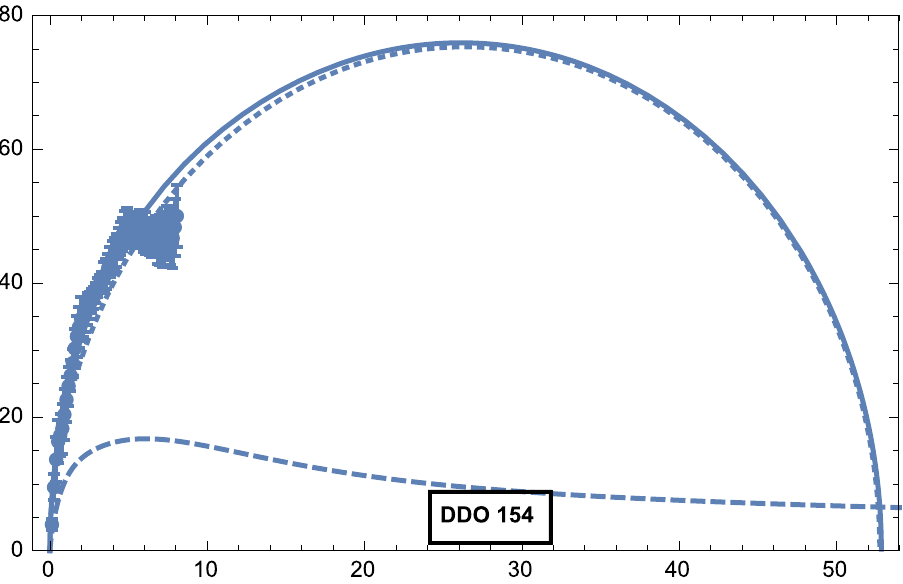, width=2.2in,height=1.1in}~~~
\epsfig{file=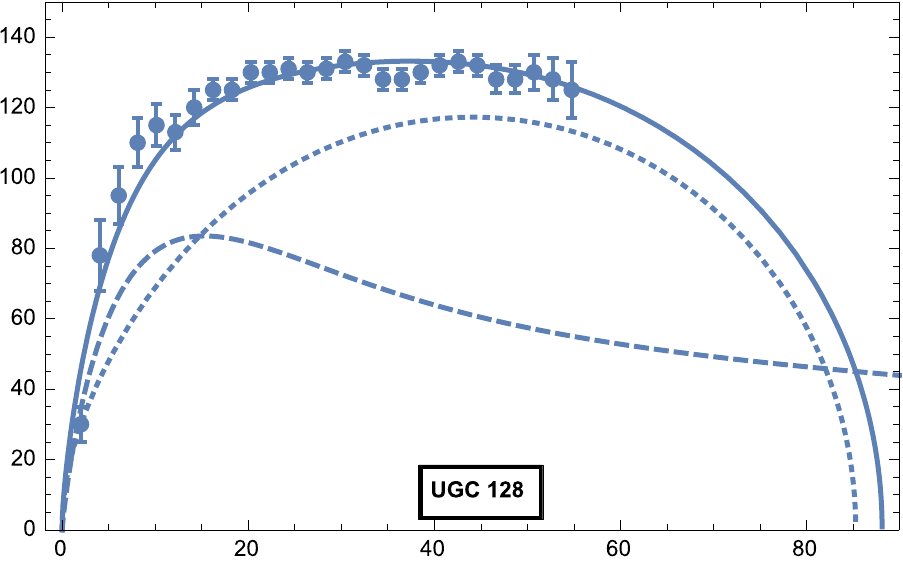, width=2.2in,height=1.1in}~~~
\epsfig{file=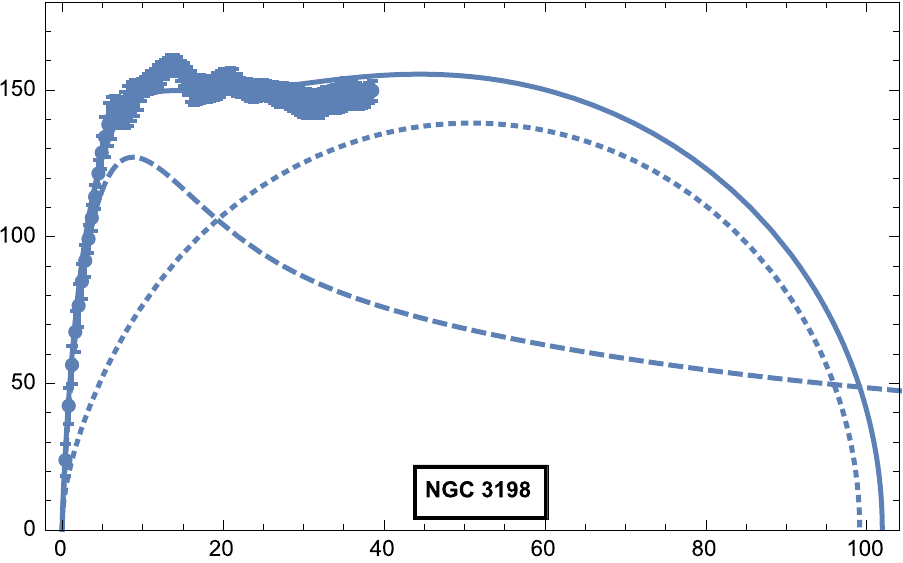, width=2.2in,height=1.1in}\\
\bigskip
\caption{Rotation velocity curve data points in ${\rm km}{\rm s}^{-1}$ versus distance $R$ in kpc for DDO 154 (dwarf spiral), UGC 128 (low surface brightness spiral), and NGC 3198 (high surface brightness spiral). The falling dashed curve is the luminous Newtonian expectation. The full curve is the conformal gravity fit, with the dotted curve being  its additional contribution. The full curve drop in velocity beyond the current data points is the conformal gravity prediction.}
\label{Fig. (9)}
\end{figure}

\section{The Newtonian solar system gold standard}

In the solar system Newton's Law of Gravity leads to orbits that to a good approximation obey $v^2/r=M_{\odot}G/r^2$. The great appeal of this law is that with one parameter, the mass $M_{\odot}$ of the Sun, one can describe the motions of all of the planets. Moreover, the Sun is a luminous source, and thus the great virtue of Newton's Law of Gravity is that once one is given the luminous matter distribution one can universally predict rotational velocities. To be able to predict velocities from luminous matter distributions should be  regarded as the Newtonian gold standard for astrophysics and gravitation, and one therefore looks to replicate it in other, typically larger, astrophysical systems. 

An additional property of Newton's Law of Gravity is that in order to determine the force at any given point on a spherical surface in a spherical matter distribution one only needs to consider the matter interior to the spherical surface. (The force falls like $1/r^2$ while the solid angle grows as $r^2$, so there is a cancellation.) Thus Newtonian gravity is local. Hence in Newtonian gravity whenever there might be a problem for which we might want to introduce more sources, we would need to put them where the problem is. For galaxies this would require putting substantial amounts of dark matter in the outskirts of galaxies where there is little visible matter, a region where the luminous Newtonian velocity expectation is falling off as $v\sim r^{-1/2}$ but the total measured velocity is not. 

There were some problems with the application of Newton's Law of Gravity to the solar system. The planet Uranus did not quite obey the law and a new planet, Neptune, one not known at the time, was proposed, with its subsequent discovery providing spectacular confirmation of Newtonian gravity. In the modern terminology Neptune should be regarded as (macho) dark matter. There was also a small but significant discrepancy for Mercury, and this was explained by the development of Einstein's General Theory of Relativity. Thus whenever there is a problem we should either introduce more sources (dark matter) or change the theory. However, the very success of Einstein's theory makes one reluctant to consider changing the theory yet again, and thus one is led to the introduction of dark matter. Nonetheless, if one were to want to make any change to Einstein's theory, one does not have to abandon the relativity principle, the covariance principle, or the equivalence principle, as one can still retain the metric as the gravitational field, and  one can still ascribe gravity to spacetime curvature. These criteria permit any general covariant gravitational action (such as the conformal gravity action we discuss below) and do not require the action to uniquely be the second-order Einstein-Hilbert action $I_{\rm EH}=-(1/16 \pi G)\int d^4x (-g)^{1/2}R^{\alpha}_{\phantom{\alpha}\alpha}$. In fact, as we show below, it is this very lack of uniqueness that leads to the dark matter problem. However, one must recover the Ricci flat Schwarzschild geometry on solar system distance scales as it has been tested there with luminous sources alone. Thus for theories that can do this the only open issue is in determining how much curvature a source might produce on larger distance scales, with a view to finding that on those larger distance scales where dark matter might be required that instead one has an alternate general relativistic pure metric theory of gravity whose luminous sources yield just the needed amount of curvature. As we see below, this is the case with the conformal gravity theory, a theory that for the purposes of this paper a reader need only consider as a foil to the standard Newton-Einstein theory.

\section{Galactic rotation curve difficulties for dark matter}

In dark matter fits dark matter is needed in regions where there is little luminous matter, and unlike the Mercury discrepancy, this is not a small effect at all. As can be seen from Fig. 1, at the last data points the ratio of the measured velocity to the luminous Newtonian expectation  in a bright spiral (NGC 3198) is of order two to one, and thus a factor of four to one in the total luminous plus dark matter potential, i.e. a dark to luminous mass ratio of order three. In a dwarf spiral (DDO 154) the velocity ratio is of order three to one, and a thus eight to one dark to luminous mass ratio. The dark to luminous ratio is thus not universal.  Bright spiral rotation curves are flat, dwarf spiral rotation curves are rising. The case where dark matter dominates (the dwarfs) is the one where rotation curves are not flat but rising, with dark matter then being needed in the inner region as well. Dwarfs thus show the dark matter problem in its starkest form. To make dark matter halo fits work for spirals one needs two free parameters per halo (e.g. the Navarro-Frenk-White  dark matter theory profile \cite{Navarro1996}). For flat rotation curves one parameter is needed to match the asymptotic value of the velocity to its maximum value in the inner region, and a second parameter is needed to keep the velocity constant in between. Also one needs two parameters per halo for dwarfs.  Thus for the 207 galaxy sample presented in \cite{Mannheim2011b,Mannheim2012c,OBrien2012,OBrien2018}, the sample on which we base our study below, one needs 414 free dark matter halo parameters. One thus looks to either derive these various parameters from first principles or to seek another theory in which no such free parameters are needed (this being the case in the conformal gravity theory discussed below).

For the moment the parameters associated with each dark matter halo are not derived from dark matter theory, and basically just parameterize the data -- you (the observer) show me the velocity and I (the dark matter theorist) will tell you the amount of dark matter. To make dark matter theory both predictive and falsifiable we require the Newtonian gold standard: you show me the luminous distribution and I will tell you the velocities. Thus dark matter theory does not a priori know how to match any given dark matter halo that it generates with any given luminous distribution. This is a serious concern because nature does know how to match velocity with luminosity (viz. visible mass), with galaxies being  found to obey the Tully-Fisher relation.

Observationally it is found that $v^4/L$ (i.e. $v^4/M$) is universal in spiral galaxies, i.e., as shown in Fig. 2, $v^4/M$ is close to universal, where $M$ is the total galactic luminous matter. Thus the velocity (due to the full gravitational potential) is correlated with the luminous distribution. Thus for dark matter to yield the Tully-Fisher relation we would need the dark to luminous ratio to be universal, but we have seen above that it is not, and thus we need to tune halo parameters galaxy by galaxy. Now in Newtonian gravity $v^2/R=MG/R^2$, so we would expect that $v^2 \sim M$. However, we find that $v^4\sim M$. So something unusual is taking place. To see what this might be let us look for other regularities that we can pull out of the 207 galaxy sample. 
\begin{figure}[H]
  \centering
 \includegraphics[width=2.2in,height=1.1in]{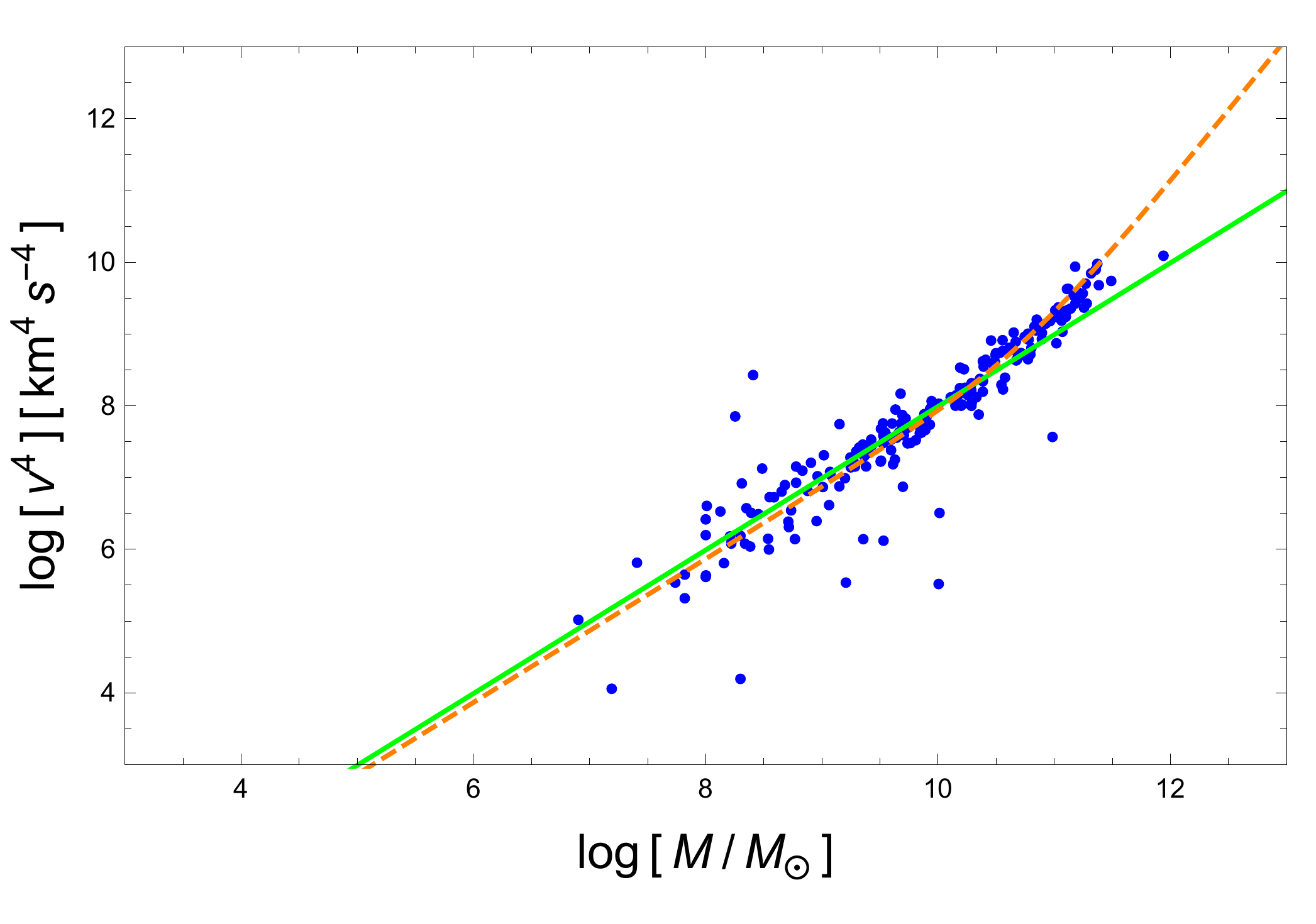}
\caption{$v^4$ versus galactic luminous mass $M$ for the last data point point in each of  a 207 spiral galaxy sample. Overlaid are $v^4=AM/M_{\odot}$ (continuous curve) and $v^4=B(M/M_{\odot})(1+N^*/D)$ (dashed curve). Here $A$ is a fitting parameter (fitted to $A=0.0098{\rm km}^4{\rm s}^{-4}$), while $B=2c^2M_{\odot}G\gamma_0=0.0074 {\rm km}^4{\rm s}^{-4}$ and $D=\gamma_0/\gamma^*=5.65\times10^{ 10}$ are fixed a priori by the conformal gravity theory.}
   \label{tullyfisher}
\end{figure}

\section{Regularities in the data}

\begin{figure}[H]
  \centering
   \includegraphics[width=2.2in,height=1.1in]{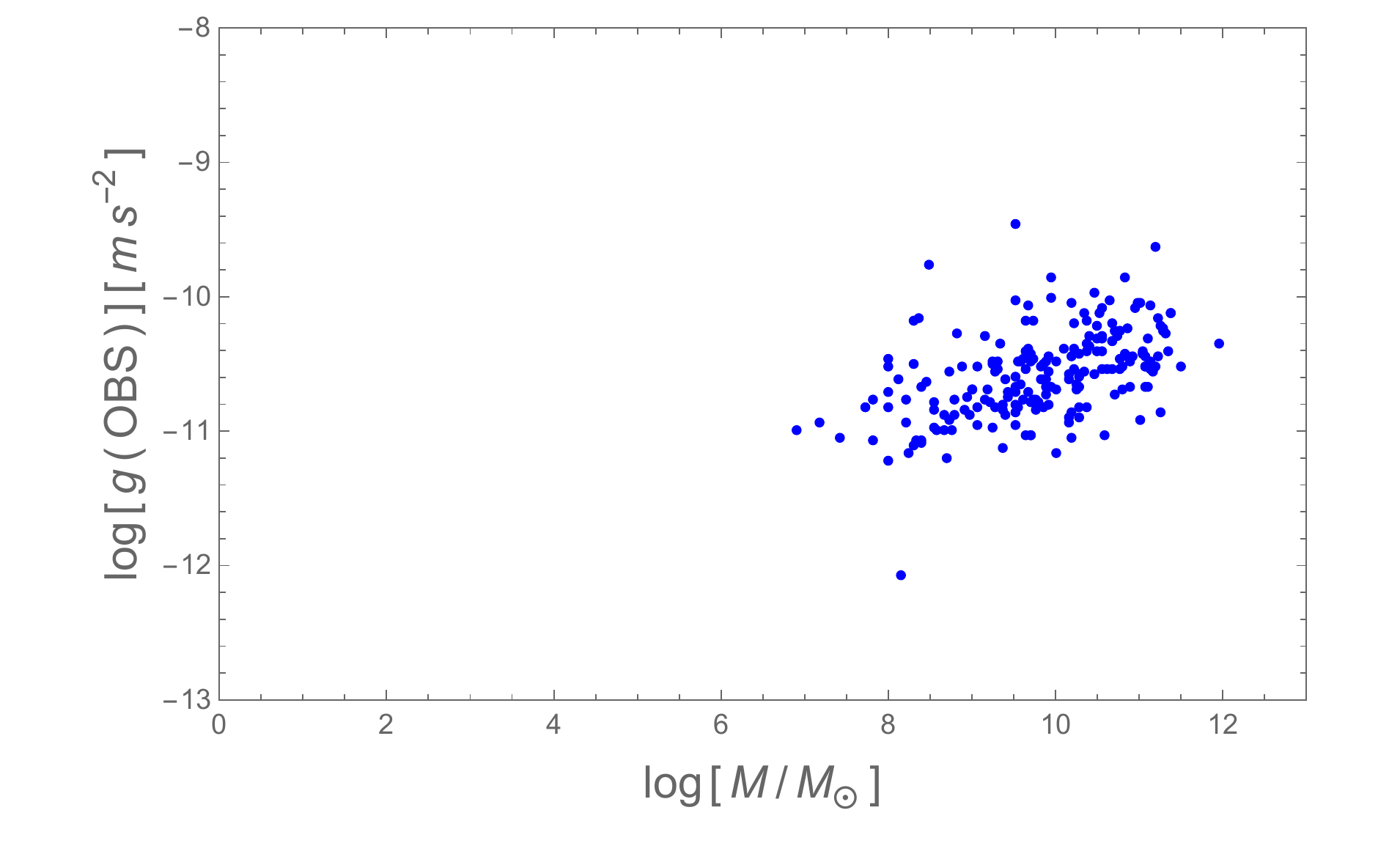}\quad  \includegraphics[width=2.2in,height=1.1in]{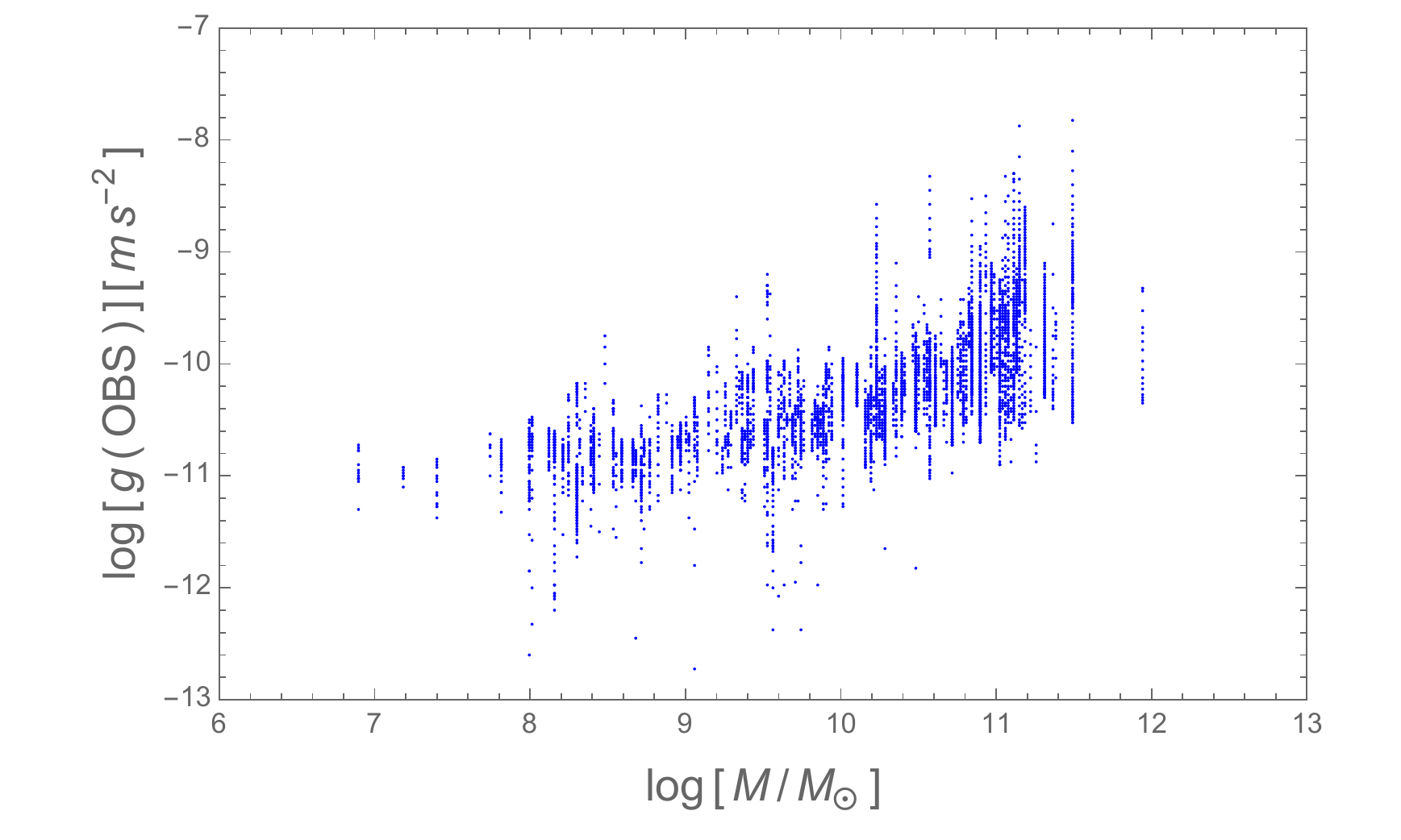}
  \caption{(a) $g({\rm OBS})$ versus $M/M_{\odot}$ for the 207 last data points, and (b) for all 5791 data points in the 207 galaxy sample.}
 \label{gobsvmall}
\end{figure}

As shown in Fig. 3(a), when plotted versus luminous galactic mass $M$, we find that the observed $g({\rm OBS})= v^2/R$ at the last data point in each galaxy is close to universal, with the data points clustering around a very small region of the plot. And not only that, numerically they cluster around a value for $v^2/R$ of order $5\times 10^{-11}{\rm m}{\rm s}^{-2}$, i.e. $v^2/c^2R$ of order $5\times 10^{-30}{\rm cm}^{-1}$, i.e. of cosmological magnitude. Moreover, as shown in Fig. 3(b), if we plot the entire 5791 data points in the 207 galaxy sample they also occupy a quite limited region of the plot.

\begin{figure}[H]
\begin{center}
 \epsfig{file=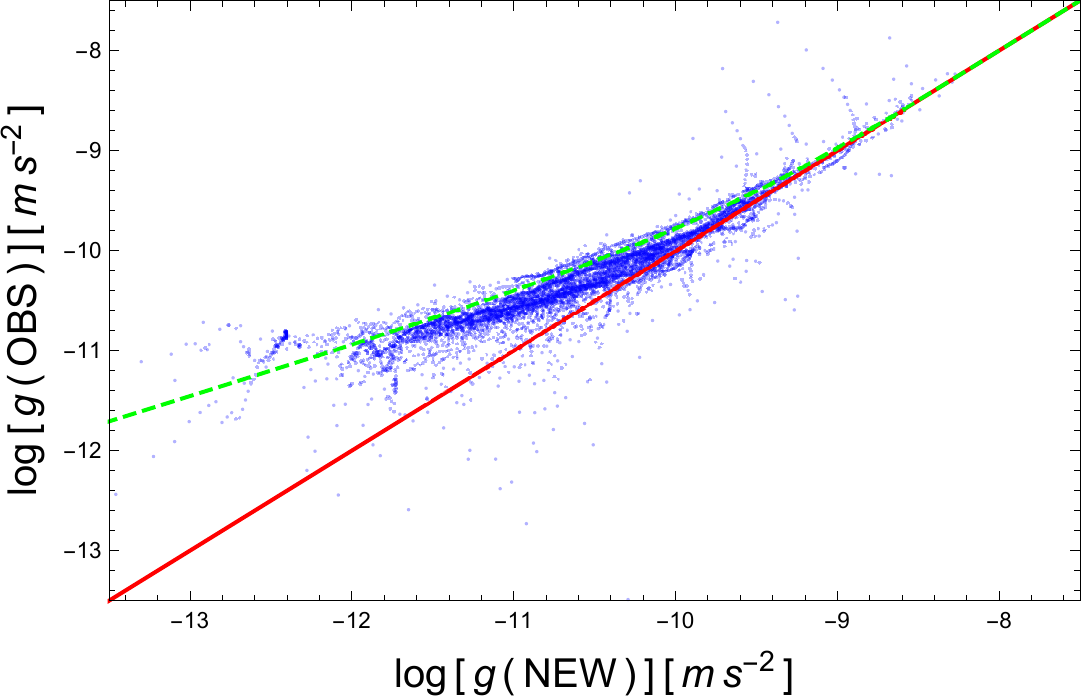,width=2.2in,height=1.1in}
 \caption{Plot of $g{\rm OBS})$ versus $g({\rm NEW})$ for the 5791 data points in our sample.  The solid  line is the line $g({\rm OBS})=g({\rm NEW})$, and the dashed line is the MLS expectation.}
 \label{dataplots}
 \end{center}
\end{figure}

To obtain an even stronger correlation we follow McGaugh, Lello and Schombert (MLS) \cite{McGaugh2016} and in Fig. 4 plot $g({\rm OBS})$ versus $g({\rm NEW})$ for the entire 5791 data points in our 207 galaxy sample, where $g({\rm NEW})$ is the Newtonian acceleration produced by the luminous matter in each galaxy. The plot is quite striking, and shows a well-defined correlation between the observed accelerations and the ones produced by the luminous matter alone. The plot shows a feature that was first noted by Milgrom in his development of MOND \cite{Milgrom1983a}, namely that  $g({\rm NEW})$ falls below $g({\rm OBS})$ for all  $g({\rm NEW})$ below a universal acceleration of order $10^{-10}~{\rm m}{\rm s}^{-2}$. To parametrize the correlation exhibited in Fig. 4 MLS introduced a phenomenological one-parameter formula
\begin{eqnarray}
g({\rm MLS})&=&\frac{g({\rm NEW})}{1-\exp(-(g({\rm NEW})/g_0)^{1/2})},
\label{FF1}
\end{eqnarray} 
with a single fundamental parameter, which we fit in Fig. 4 to the value   $g_0=6\times 10^{-11}~{\rm m~s}^{-2}$. Apart from actually exhibiting a correlation between $g({\rm OBS})$ and $g({\rm NEW})$, the utility of Fig. 4 is that it takes the rotation curves such as those in Fig. 1 and puts all the 5791 data points in our sample on one single plot. As such Fig. 4 meets the Newtonian gold standard, namely given a luminous $g({\rm NEW})$ alone one can read off $g({\rm OBS})$. We shall see below that there will be some additional considerations at very low $g({\rm NEW})$ where $g({\rm MLS})\rightarrow g_0^{1/2}g({\rm NEW})^{1/2}$, but regardless of this, Fig. 4 does not immediately point in the direction of dark matter. Fig. 4 does demand an explanation, and as we now show, the conformal gravity theory provides one.

\section{Conformal gravity fits}

The conformal gravity theory \cite{Mannheim1989,Mannheim1994,Mannheim2006,Mannheim2012b,Mannheim2017} is based on the gravitational action 
\begin{eqnarray}
I_W&=&-\alpha_g\int d^4x (-g)^{1/2}C_{\lambda\mu\nu\kappa} C^{\lambda\mu\nu\kappa}
=-2\alpha_g\int d^4x (-g)^{1/2}\left[R^{\mu\nu}R_{\mu\nu}-{1 \over 3}(R^{\alpha}_{\phantom{\alpha}\alpha})^2\right],
\label{FF2}
\end{eqnarray}
where $\alpha_g$ is a dimensionless gravitational coupling constant. For a static, spherically symmetric source with energy-momentum tensor $T_{\mu\nu}$, the conformal gravity equations of motion reduce without approximation to the fourth-order Poisson equation \cite{Mannheim1994}
\begin{eqnarray}
\nabla^4B=\frac{3}{4\alpha_gB(r)}(T^{0}_{\phantom{0}0}-T^{r}_{\phantom{r}r}),
\label{FF3}
\end{eqnarray}
where $B(r)=-g_{00}=1/g_{rr}$. For a star localized to a region $r<r_0$  the metric and gravitational potential produced by it in the $r>r_0$ region are given by \cite{Mannheim1989,Mannheim1994}
\begin{eqnarray}
B(r)=-g_{00}=\frac{1}{g_{rr}}=1-\frac{2\beta^*}{r}+\gamma^*r,\quad V^{*}(r)=-\frac{\beta^{*}c^2}{r}+\frac{\gamma^{*} c^2 r}{2},
\label{FF4}
\end{eqnarray}
to thus generalize both the Newtonian potential and the Schwarzschild solution to Einstein gravity, with departures from them only occurring at the large distances where dark matter is ordinarily required.

For a spiral galaxy with $N^*$ stars, galactic mass $M=N^*M_{\odot}$, and surface brightness $\Sigma (R)=\Sigma_0e^{-R/R_0}$, we obtain the local contribution due to all the stars in the galaxy of the form
\begin{eqnarray}
\frac{v_{{\rm LOC}}^2}{R}&=&
\frac{N^*\beta^*c^2 R}{2R_0^3}\left[I_0\left(\frac{R}{2R_0}
\right)K_0\left(\frac{R}{2R_0}\right)-
I_1\left(\frac{R}{2R_0}\right)
K_1\left(\frac{R}{2R_0}\right)\right]
\nonumber \\
&&+\frac{N^*\gamma^* c^2R}{2R_0}I_1\left(\frac{R}{2R_0}\right)
K_1\left(\frac{R}{2R_0}\right) \rightarrow \frac{N^*\beta^*c^2}{R^2}+
\frac{N^*\gamma^*c^2}{2},
\label{FF5}
\end{eqnarray} 
with the indicated large $R$ behavior.

If this were to be the whole story, we would then obtain $v^2\sim M$ and fail to account for the Tully-Fisher $v^4\sim M$ regularity observed in spiral galaxies. However since the force is not falling as $1/R^2$ while the solid angle is still growing as $R^2$, Newton's theorem that we can exclude the gravitational force due to exterior matter no longer applies. Consequently, unlike Newtonian gravity, conformal gravity is global, with a test particle in a galaxy seeing both the local interior galactic field and the global exterior gravitational field, viz. the rest of the visible Universe. Moreover, since the potential is growing with distance the matter furthest away from any given galaxy would be the most important, viz. cosmology itself, to thus give an effect that is both of cosmological origin and independent of the mass of any given galaxy, to thus be universal. Cosmology actually supplies two contributions, one due to the Hubble flow and the other due to the inhomogeneities in it. 

Explicit calculation shows that the Hubble flow yields a universal linear potential \cite{Mannheim1989} (with parameter $\gamma_0$) while the inhomogeneities yield a universal quadratic potential \cite{Mannheim2011b} (with parameter $\kappa$), to give a conformal gravity (CG) total velocity $v({\rm TOT})$ that obeys

\begin{equation}
g({\rm CG})=\frac{v^2_{\rm TOT}}{R}=\frac{v^2_{\rm LOC}}{R}+\frac{\gamma_0 c^2}{2}-\kappa c^2R
\rightarrow \frac{N^*\beta^*c^2}{R^2}+
\frac{N^*\gamma^*c^2}{2}+\frac{\gamma_0c^2}{2}-\kappa c^2R.
\label{FF6}
\end{equation} 

\medskip
\noindent
Conformal gravity has now successfully fitted 207 galaxies  \cite{Mannheim2011b,Mannheim2012c,OBrien2012,OBrien2018}, with a typical three of them being shown in Fig. 1. In the fits the visible $N^*$ of each galaxy is the only variable, with fitted $\beta^* = M_{\odot}G/c^2 = 1.48 \times 10^5{\rm cm}$, $\gamma^* = 5.42 \times 10^{-41} {\rm cm}^{-1}$, $\gamma_0 = 3.06 \times 10^{-30}{\rm  cm}^{-1}$, and $\kappa = 9.54 \times 10^{-54} {\rm cm}^{-2}$ all being universal, with no dark matter being needed, and with 414 fewer free parameters than in dark matter calculations. Now the reader might be concerned with the notion of potentials that grow with distance, but as we see in Fig. 1, the piece that one needs to add on to the luminous Newtonian contribution is not itself flat but is actually rising until the last data point. And as we see from the structure of $g({\rm CG})$, it is actually rising universally. In the fitting $\gamma_0$ is not just universal but is naturally found to be a cosmological scale (to thus provide a natural origin for a universal acceleration of the type suggested by Milgrom), while $\kappa$ is naturally found to be a cluster of galaxy scale; with both $\gamma_0$ and $\kappa$  precisely being found to be of the cosmological background and inhomogeneity scales that led to their presence in the first place. In the view of conformal gravity dark matter is just an attempt to describe this global physics in local terms, and is thus not required to exist. In the view of conformal gravity missing mass actually is luminous, being not in galaxies at all but being due to the visible matter in the rest of the Universe. In other words, the missing mass has been hiding in plain sight all along.

\section{Conformal gravity fitting to the $g({\rm OBS})$ versus $g({\rm NEW})$ plot}

To show that conformal gravity does fit all the data, following 
\cite{OBrien2018} (where the figures presented in this paper may be found) we overlay $g({\rm CG})$ on a plot of $g({\rm OBS})$ versus $g({\rm NEW})$ for the 207 galaxies. As we see in Fig. 5, conformal gravity precisely lines up with the data points, doing so by fitting the full width and not just yielding a single curve through the data. The width observed  in Fig. 5 is thus physical.
\begin{figure}[H]
  \centering
    \includegraphics[width=2.2in,height=1.1in]{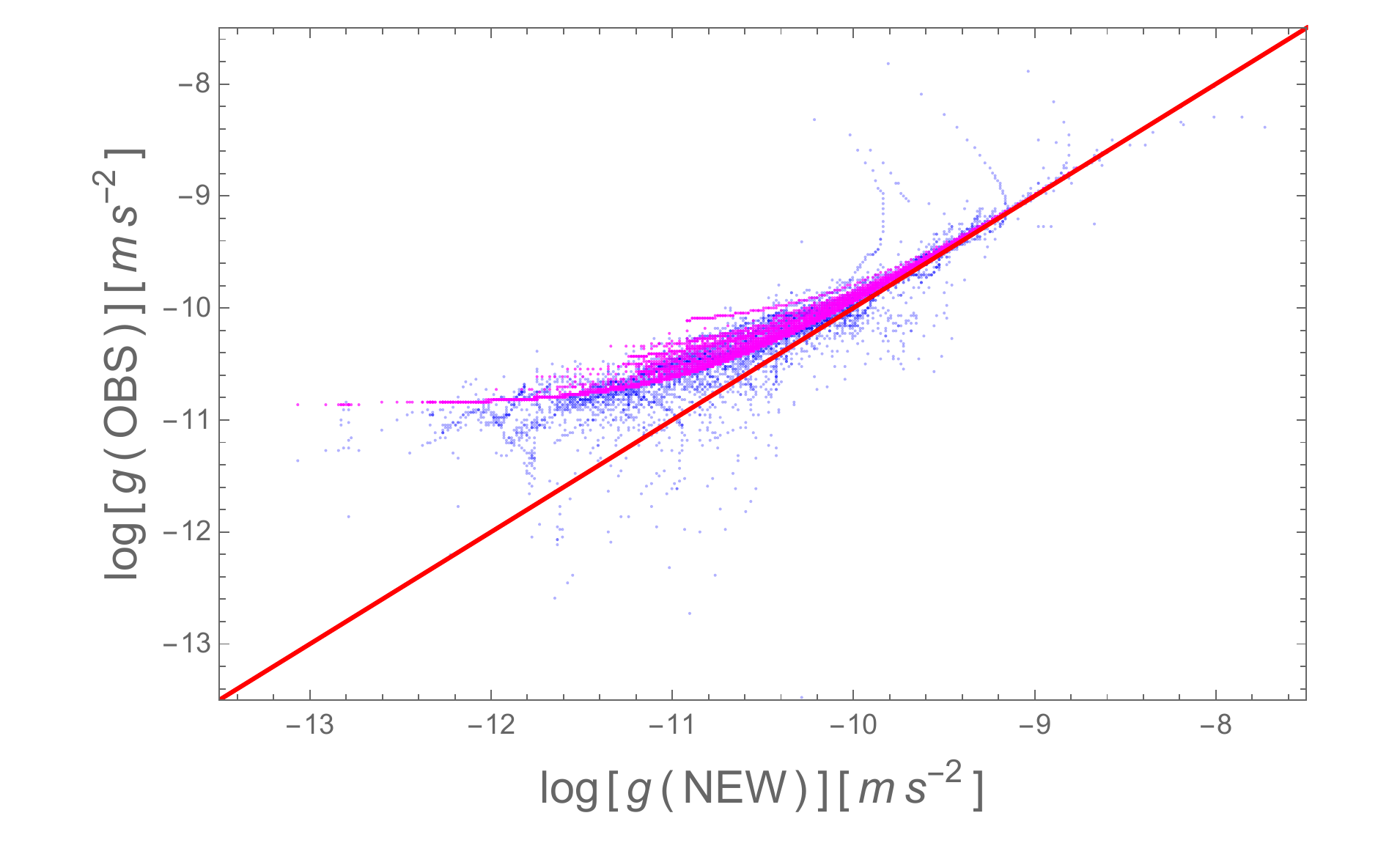}
  \caption{$g({\rm CG})$ overlay of $g({\rm OBS})$ versus $g({\rm NEW})$. The solid lines other than the  $g({\rm OBS})=g({\rm NEW})$ diagonal are the $g({\rm CG})$ expectations. }
   \label{cgoverlayall}
\end{figure}

\noindent
To study the structure of the fit in more detail, in Fig. 6(a) we overlay $g({\rm CG})$ on a plot of $g({\rm OBS})$ versus $g({\rm NEW})$ for  the 2870 data points in the 56 HSB (high surface brightness) galaxies in our 207 galaxy sample. Similarly, in Fig. 6(b) we overlay $g({\rm CG})$ on a plot of $g({\rm OBS})$ versus $g({\rm NEW})$ for  the 2921 data points in the 151 LSB (low surface brightness and dwarf) galaxies in our 207 galaxy sample. As we see, it is just the HSB galaxies that fill out the width of the plot. For the LSB sample there is just a single curve, one with a very interesting continuation to very small $g({\rm NEW})$, namely unlike the $g({\rm MLS})$ curve discussed above, the conformal gravity curve is flattening off at very low $g({\rm NEW})$ and becoming independent of $g({\rm NEW})$ altogether. This flattening is to be expected since until the $\kappa$ term  becomes important $g({\rm CG})$ is asymptoting to $\gamma_0c^2/2$, a pure constant. (At large enough $R$ where the $\kappa$ term does become important, $v^2$ is predicted to fall, just as shown in Fig. 1. However, since $v^2$ cannot go negative, galaxies would have to end. Conformal gravity thus requires that no galaxy could be bigger than a maximum size (this was first heuristically suggested in \cite{Kazanas1991}), one numerically of order 150 kpc.)

\begin{figure}[H]
  \centering
    \includegraphics[width=2.2in,height=1.1in]{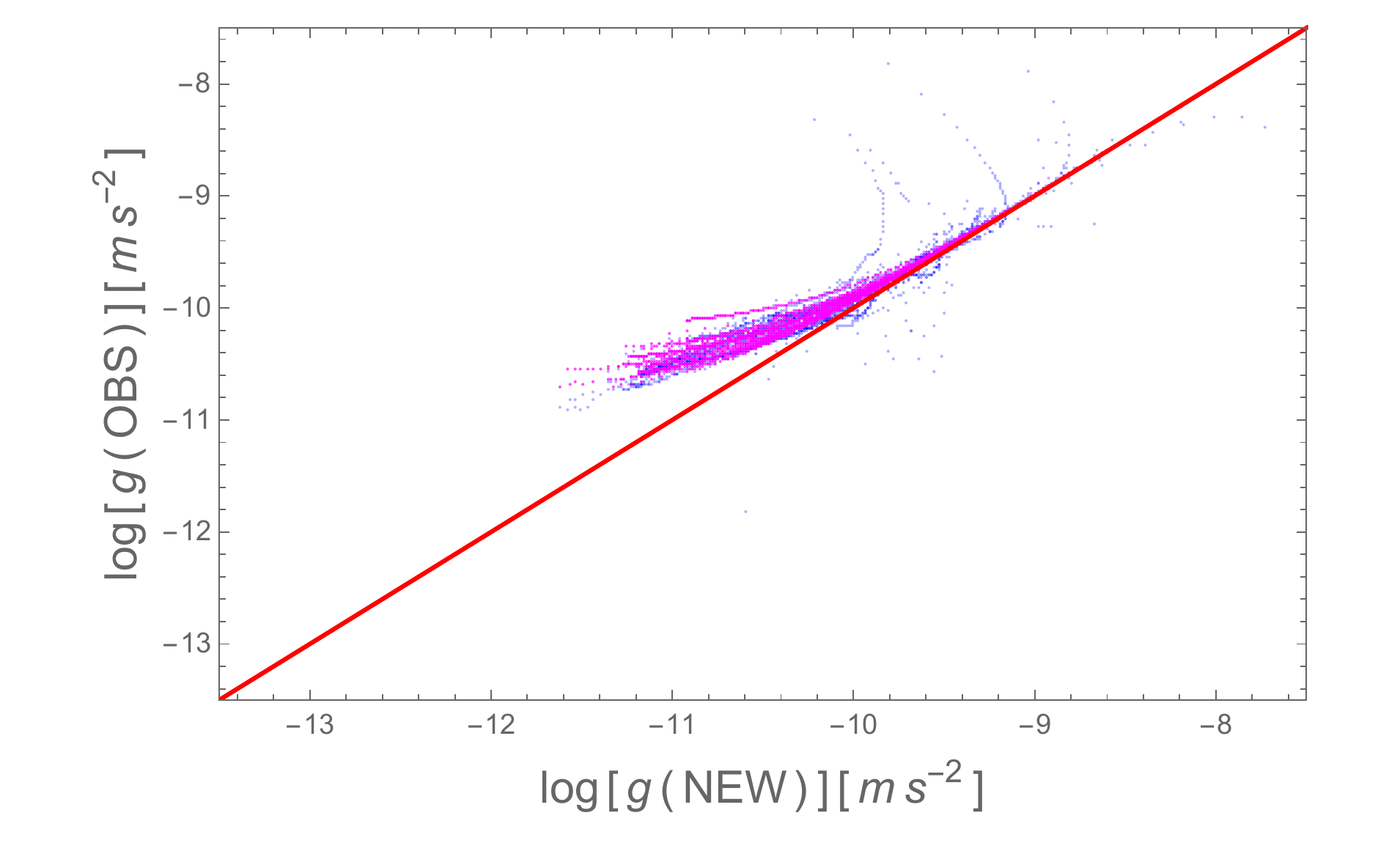}\quad \includegraphics[width=2.2in,height=1.1in]{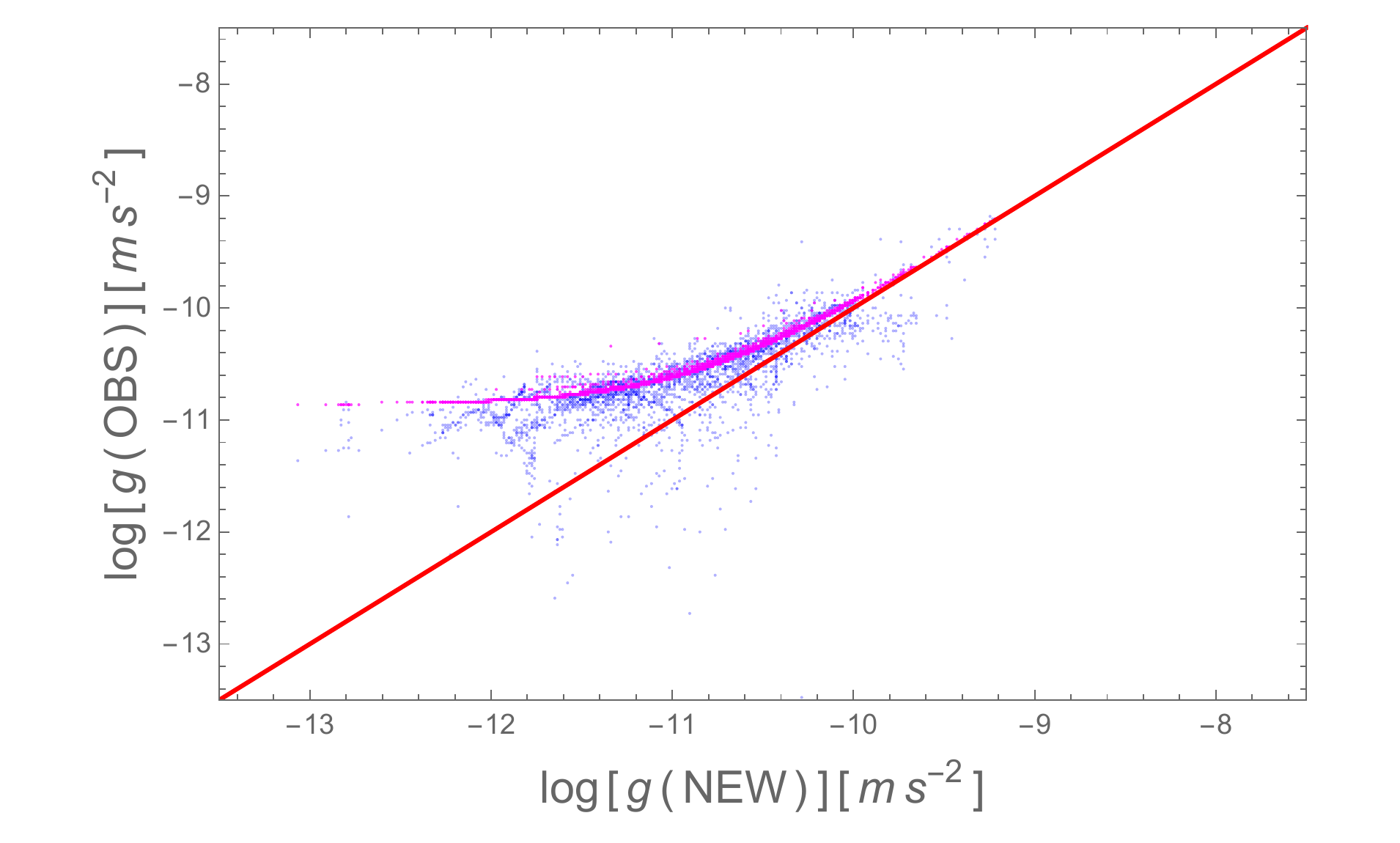}
  \caption{$g({\rm CG})$ overlay of  $g({\rm OBS})$ versus  $g({\rm NEW})$ for (a) HSB and (b) LSB galaxies. The lines other than the  $g({\rm OBS})=g({\rm NEW})$ diagonal are the $g({\rm CG})$ expectations.}
   \label{cgoverlayhsb}
\end{figure}

\noindent
To study this very small $g({\rm NEW})$ region  Lelli, McGaugh, Schombert, and Pawloski \cite{Lelli2017} augmented the spiral galaxy data with some dwarf spheroidal data and some late type galaxy data, and found that there is in fact a flattening off at very low $g({\rm NEW})$. They even considered changing the $g(MLS)$ formula by adding on a term  $\hat{g}\exp(-g({\rm NEW})g_0/\hat{g}^2)^{1/2}$ where $\hat{g}$  is a new free parameter, and they characterized the data as exhibiting a possible ``acceleration floor". In conformal gravity such an acceleration floor is natural.

\section{A distance-dependent regularity}
\begin{figure}[H]
  \centering
    \includegraphics[width=2.2in,height=1.1in]{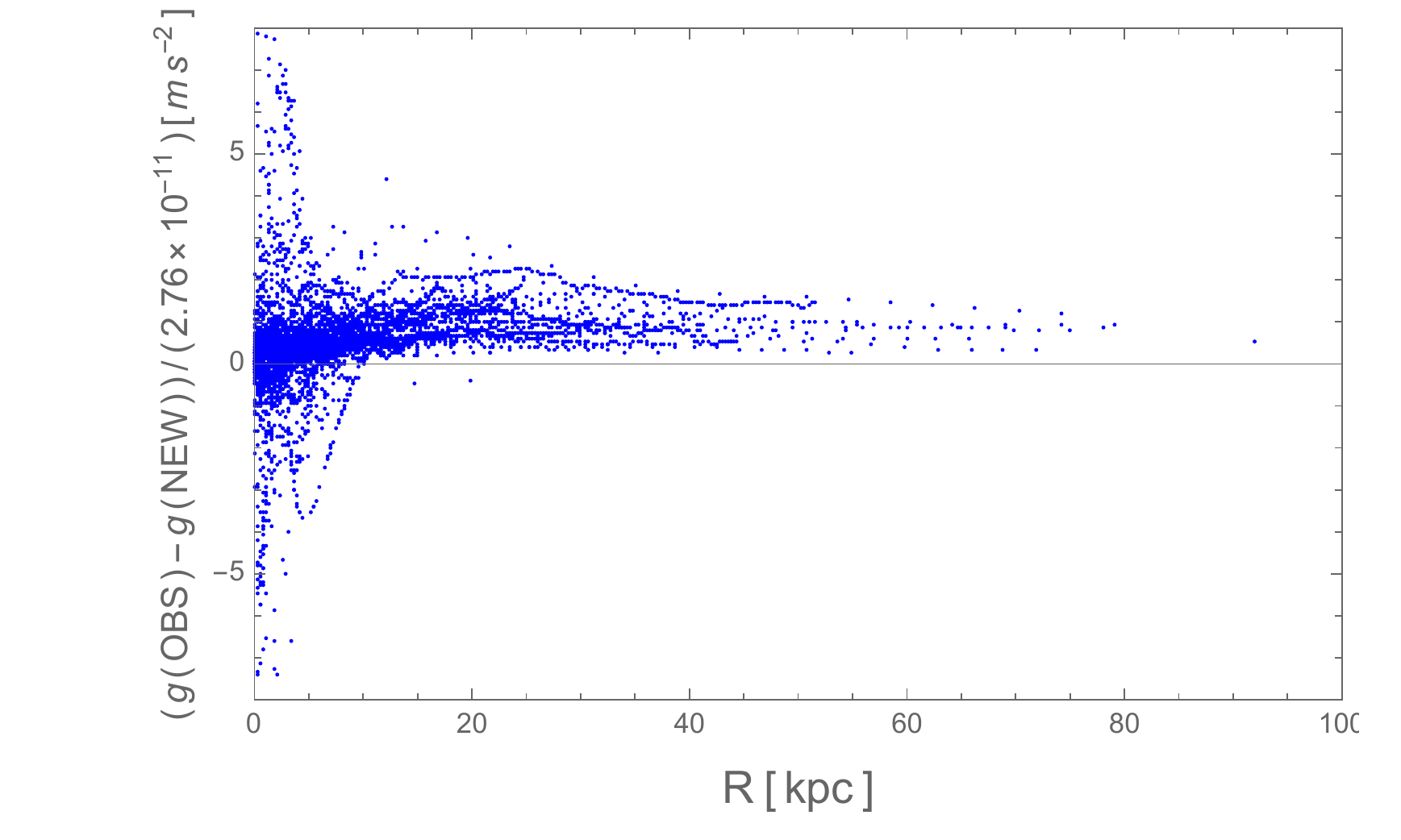}
  \caption{$g({\rm OBS})-g({\rm NEW})$ as scaled by  $1/(2.76\times 10^{-11})$ versus $R$.}
   \label{versusr}
\end{figure}

\noindent
While plotting $g({\rm OBS})$ versus $g({NEW})$  is instructive, using $g({\rm NEW})$ as a parameter is somewhat non-intuitive. However, it is possible to make a plot involving a more intuitive parameter, namely distance. Thus in Fig. 7 we plot $g({\rm OBS})-g({\rm NEW})$ (as scaled by $1/\gamma_0c^2=1/(2.76\times 10^{-11}{\rm ms}^{-2})$) versus $R$. As we see, by 10 kpc we get an increase over the luminous Newtonian expectation in every data point in every one of the 207 galaxies. And not only that, the shortfall between $g({\rm NEW})$ and $g({\rm OBS})$ is almost constant in distance, just as expected of $g({\rm CG})$.

\section{Tully-Fisher relation}
\medskip
\noindent
\noindent
With  conformal gravity yielding $v^2=\beta^*c^2N^*/R+(\gamma^*N^*+\gamma_0)c^2R/2$ in the intermediate region in a galactic rotation curve, at the crossing point $R_c$ between the $1/r$ and $r$ conformal gravity potential contributions one can set $\beta^*c^2N^*/R_c=(\gamma^*N^*+\gamma_0)c^2R_c/2$ for an $R_c$ that depends on each galaxy. And thus at that point one can set $v^4=B(M/M_{\odot})(1+N^*/D)$ where $B=2c^2M_{\odot}G\gamma_0=0.0074$ km$^4$s$^{-4}$, where $D=\gamma_0/\gamma^*=5.65\times10^{10}$, and where here $M=N^*M_{\odot}$ includes all galactic baryonic sources. Since the velocities at the last data points do not differ much from those at the crossover points in each galaxy, at the last data points we plot $v^4=B(M/M_{\odot})(1+N^*/D)$ as the dashed curve in the Fig. 1, to thus establish not only that conformal gravity naturally leads to the Tully-Fisher relation, it actually determines the coefficients that are involved. In addition, the theory naturally suggests a typical largest value for $N^*$, viz. the  value $N^*=\gamma_0/\gamma^*=5.65\times10^{10}$ for which the locally generated $\gamma^*N^*$ and the globally generated $\gamma_0$ terms  balance each other. Interestingly, in a plot of the density distribution of galaxies versus galactic stellar  mass (the galaxy stellar mass function), a sharp drop in density sets in at close to this value, with there being very few  galaxies with stellar mass larger than $5.65\times10^{10}M_{\odot}$ or so \cite{Baldry2008}. Moreover, at the value $\gamma^*N^*=\gamma_0$  the crossing point obeys $R_c^2=\beta^*N^*/\gamma_0=MG/c^2\gamma_0$, to yield the mass-radius relation that was heuristically  suggested in \cite{Kazanas1991}.

\section{Lack of necessity of the second-order Poisson equation or the Einstein equations}

The second-order, fourth-order, and sixth-order Poisson equations have exterior solutions of the form
\begin{eqnarray}
\nabla^2\phi&=&\rho,~\phi=-\frac{\beta}{r};~~~~~\qquad
\nabla^4\phi=\rho,~\phi=-\frac{\beta}{r}+\gamma r;~~~~~\qquad\nabla^6\phi=\rho,~\phi=-\frac{\beta}{r}+\gamma r +\delta r^3.
\label{FF7}
\end{eqnarray} 
As we see, in each case we obtain a $1/r$ solution. The argument generalizes to arbitrary even order. The second-order Poisson equation is thus sufficient to give Newton's Law of Gravity, but not necessary. It is assuming it to be necessary that creates the dark matter problem. While the higher-derivative Poisson equations also lead to a $1/r$ potential they only modify it at large distances, viz. just where dark matter is needed. Also we note that while the $\phi=-\beta /r +\gamma r$ solution to $\nabla^4\phi=\rho$ does reduce to the $\phi=-\beta /r $ solution to $\nabla^2\phi=\rho$ at small enough $r$,  $\nabla^4\phi=\rho$ itself does not reduce to  $\nabla^2\phi=\rho$. Thus we do not need higher-order derivative equations of motion to reduce to  second-order derivative equations of motion in order to get the higher-derivative solutions to reduce to solutions to the second-order equations. Since observationally one only ever needs to recover the solutions in the region in which they have been tested (i.e. tested without needing to invoke dark matter), one can bypass the second-order Poisson equation, and thus by extension the Einstein equations that produce it, altogether.

Relativistically, the Einstein equations reduce to $\nabla^2\phi=\rho$, while the conformal gravity equations reduce to $\nabla^4\phi=\rho$. Moreover, while the Einstein equations lead to the Schwarzschild solution, the Schwarzschild solution is also an exterior solution to the  conformal theory. The Einstein equations are thus sufficient to give the Schwarzschild solution, but not necessary. It is the assumption that the Einstein equations are necessary that leads to the dark matter problem. 

Thus we need a principle in order to determine whether any particular theory of gravity might also be necessary and not just sufficient. Local conformal invariance (i.e. invariance under $g_{\mu\nu}(x)\rightarrow e^{2\alpha(x)}g_{\mu\nu}(x)$ for arbitrary $\alpha(x)$) supplies such a principle since conformal gravity is the only theory that one can write down in four spacetime dimensions that obeys it. Moreover, not only does conformal gravity eliminate the need for galactic dark matter,  it also tames the cosmological constant while providing for a cosmology that fits the accelerating Universe data naturally without needing any fine tuning or needing any dark matter \cite{Mannheim2017b}. And moreover, the conformal theory is even renormalizable and unitary and ghost free at the quantum gravity level \cite{Bender2008a}. While more work remains to be done on the conformal gravity theory, it does well on the testing that has so far been applied to it, and its success in accounting for galactic rotation curve systematics with universal parameters challenges dark matter theory to do so too. So even if the conformal gravity theory might not be valid, its formulae are valid as they do account for the data, and thus they and the regularities that we have presented in this paper require that dark matter theory reproduce them.

\end{document}